\documentclass[%
 aip,
 jmp,%
 amsmath,amssymb,
 reprint,%
]{revtex4-1}

\usepackage{graphicx}
\usepackage{dcolumn}
\usepackage{bm}
\begin{document}

\title[Interacting Dark Energy]{Models of Interacting Dark Energy}

\author{W. Zimdahl}

\affiliation{Departamento de
F\'{\i}sica, Universidade Federal do Esp\'{\i}rito Santo\\
Vit\'{o}ria, Esp\'{\i}rito Santo, Brazil}

\date{\today}
\begin{abstract}
Any non-gravitational coupling between dark matter and dark energy modifies the cosmological dynamics.
Interactions in the dark sector are considered to be relevant to address the coincidence problem. Moreover, in various models the observed accelerated expansion of the Universe is a pure interaction phenomenon. Here we review recent approaches in which a coupling between both dark components is crucial for the evolution of the Universe.
\end{abstract}



\maketitle

\section{\label{sec:introduction}Introduction}
According to the prevailing view, our present Universe is dynamically dominated two so far unknown components, dark matter (DM) with an energy density $\rho_{m}$ and dark energy (DE) with a density $\rho_{x}$. DM is usually modeled as a pressureless fluid, while DE is characterized by an equation of state (EoS) parameter $w$. The determination of this parameters is one of the major challenges in the field. Current observations are consistent with a spatially flat universe with fractions of 72\% DE and 28\% matter (including DM and baryons). The preferred model is the $\Lambda$CDM model
 ($\Lambda$ denotes the cosmological constant, equivalent to a constant energy density
 $\rho_{x} = \rho_{\Lambda}$ and ``CDM" stands for cold dark matter) for which $w = -1$. Because of the cosmological constant problem in its different facets, competing approaches were developed which either ``dynamize" the cosmological constant or modify General Relativity.
 Here we review various approaches in which DM and DE do not evolve separately but interact with each other non-gravitationally. This gives rise to a richer cosmological dynamics compared with non-interacting models. Moreover, disregarding the potential existence of a coupling between both dark components may result in a misled interpretation of observational data\cite{das}.

\section{The dark sector of the Universe}

\subsection{The $\Lambda$CDM model}
The basic equations of the $\Lambda$CDM model are
\begin{equation}\label{friedmann}
H^{2}
= \frac{8 \pi G}{3} \rho_{m} -
\frac{k}{a ^{2}} + \frac{\Lambda}{3}
\end{equation}
and
\begin{equation}\label{dda}
\frac{\ddot{a}}{a} = - 4 \pi G\left(\rho_{m} + 3 p_{m} \right) +
\frac{\Lambda}{3}\ ,
\end{equation}
where $H = \frac{\dot{a}}{a}$ is the Hubble rate and $a$ is the scale factor of the Robertson-Walker metric.

Introducing the critical density $\rho_{c}\equiv \frac{3 H^{2}}{8\pi G}$ and the fractional quantities
\begin{equation}\label{frac}
\Omega_{m} \equiv \frac{\rho_{m}}{\rho_{c}}\ , \
 \Omega_{\Lambda } \equiv \frac{\Lambda }{3H ^{2}}\ ,\
 \Omega_{k} \equiv -\frac{k}{a ^{2}H ^{2}}\ ,
\end{equation}
the
Friedmann equation (\ref{friedmann}) can be rewritten as
\begin{equation}\label{sum}
\quad\Omega _{m}
+ \Omega _{k} + \Omega _{\Lambda } = 1\ .
\end{equation}
The current data are consistent with (the subindex $0$ denotes present values)
\begin{equation}\label{current}
\Omega _{m0} \approx 0.28,  \ \ \Omega _{\Lambda0} \approx 0.72
\quad\Rightarrow\quad \Omega _{k0}\approx 0\ .
\end{equation}
With the definitions $\Lambda \equiv 8\pi\,G\,\rho_{\Lambda}$ and
$p_\Lambda \equiv - \rho_{\Lambda}$, the  cosmological constant
can be seen  as a ``fluid" with negative
pressure. Then
\begin{equation}\label{lnegp}
H^{2} = \frac{8 \pi G}{3}\rho - \frac{k}{a ^{2}}
\ ,\
\
\frac{\ddot{a}}{a} = - \frac{4 \pi G}{3}\left(\rho + 3 p\right) \ ,
\end{equation}
where $\rho = \rho_{m } + \rho_{\Lambda}$ and $p = p_{m } + p_{\Lambda}$.
On the other hand, the cosmological constant is dynamically equivalent to the lowest energy state of a scalar field with energy-momentum tensor
\begin{equation}\label{Tscalar}
T_{mn} = \varphi_{,m}\varphi_{,n} -
g_{mn}\left(\frac{1}{2}g^{ik}\varphi_{,i}\varphi_{,k} +
V(\varphi)\right)\ .
\end{equation}
Namely, considering the configuration
\begin{equation}\label{Tvac}
T_{mn}^{vac} =  - g_{mn} V(\varphi_{min}) \equiv -
\rho_{vac}g_{mn}\ ,
\end{equation}
it becomes obvious, that in the
field equations
\begin{equation}\label{Einstein}
R ^{ik}- \frac{1}{2}\,g ^{ik}\,R + \Lambda g_{mn} =
8\pi\,G\,T_{mn}
\end{equation}
there appears an
effective cosmological
``constant" $\Lambda_{eff} = \Lambda + 8\,\pi\,G\,\rho_{vac}$.
The density $\rho_{vac}$ of the vacuum energy (recall $E_{vac} = \frac{1}{2}\sum\hbar \omega$ and $\omega = \sqrt{k^{2} + m^{2}}$) is calculated via
\begin{equation}\label{rhovac}
\rho_{vac} =
\frac{1}{2}\,\frac{1}{\left(2\,\pi\right)^{3}}\int_{0}^{k_{c}}4\,\pi\,k^{2}k\mbox{d}k
= \frac{k_{c}^{4}}{16\,\pi^{2}}\ ,
\end{equation}
where we have assumed $\hbar = 1$ and $m = 0$.
For a cutoff $k_{c}$ at the Planck scale one finds
$\rho_{vac} \sim 10^{76} {\rm
(GeV)}^{4}\ \approx 10^{114}
\frac{\mathrm{erg}}{\mathrm{cm}^{3}}$.
The observed value for the effective cosmological ``constant" is
$\rho_{\Lambda_{eff}} \sim 10^{-47} {\rm
(GeV)}^{4}\approx 10^{-9}
\frac{\mathrm{erg}}{\mathrm{cm}^{3}}$.
This large discrepancy of 123 orders of magnitude constitutes the cosmological constant problem.

Another problem results from the circumstance that the ratio of the energy densities,
$r = \frac{\rho_{m}}{\rho_{x}}$, has a present value of  $r_{0} \approx \frac{3}{7} $ which is of the order of unity.
If one takes into account that, for the $\Lambda$CDM model, $\rho_{x} = \rho_{\Lambda} =$ const and $\rho_{m} \propto a^{-3} $, it follows that
$r \propto a^{-3}$. A value of  $r_{0}$ of the order of one seems to single out a very special period. In general, $r$ differs from unity by many orders of magnitude. This is called the
coincidence problem.

\section{Interacting dark energy}

In the $\Lambda$CDM model and in most other approaches DM and DE are considered to be uncoupled components with separately conserved energy-momentum balances. However, Einstein's equations only require total energy-momentum conservation. This freedom can be used to construct alternative models of the dark sector in which DM and DE also interact non-gravitationally. We start with some general relations before considering specific models. We assume the total energy-momentum tensor to have a perfect fluid structure,
$T_{ik} = \rho u_{i}u_{k} + p h_{ik}$ with  $T_{\ ;k}^{ik} = 0$. Further, we assume a split into two components,
\begin{equation}\label{split}
T^{ik} = T_{m}^{ik} + T_{x}^{ik}
\end{equation}
with perfect-fluid structures as well, i.e.,
$T_{A}^{ik} = \rho_{A} u_A^{i} u^{k}_{A} + p_{A} h_{A}^{ik}$ and $ h_{A}^{ik} = g^{ik} + u_A^{i} u^{k}_{A}$, where  $A = m, x$.
A coupling of both components is described by
\begin{equation}\label{coupling}
T_{m\ ;k}^{ik} = Q^{i},\qquad T_{x\ ;k}^{ik} = - Q^{i}\ ,
\end{equation}
which guarantees the overall energy-momentum conservation.
The separate energy balance equations are
\begin{equation}\label{ebalm}
\rho_{m,a}u_{m}^{a} +  \Theta_{m} \left(\rho_{m} + p_{m}\right) = -u_{ma}Q^{a}
\end{equation}
and
\begin{equation}\label{ebalx}
\rho_{x,a}u_{x}^{a} +  \Theta_{x} \left(\rho_{x} + p_{x}\right) = u_{xa}Q^{a}\ ,
\end{equation}
where $\Theta_{A} \equiv u^{k}_{A;k}$.
Likewise, we have the momentum balances
\begin{equation}\label{mbalm}
\left(\rho_{m} + p_{m}\right)\dot{u}_{m}^{a} + p_{m,i}h_{m}^{ai} = h_{m i}^{a} Q^{i}
\end{equation}
and
\begin{equation}\label{mbalx}
\left(\rho_{x} + p_{x}\right)\dot{u}_{x}^{a} + p_{x,i}h_{x}^{ai} = - h_{x i}^{a} Q^{i}\ .
\end{equation}
Generally, the four velocities of the components are different from each other and from the total four velocity. For the homogeneous and isotropic background, however, we assume them to coincide:
$u_{m}^{a} = u_{x}^{a} = u^{a}$. Restricting ourselves for the moment to the background, the individual energy balances are
\begin{equation}\label{balfundo}
\dot{\rho}_{m} + 3H \rho_{m} = Q \ ,\quad \dot{\rho}_{x} +
3H (1+w)\rho_{x} = - Q \ ,
\end{equation}
where $Q = - u_{a}Q^{a}$ is considered to be a phenomenological quantity. Obviously, $Q$ does not directly enter the Hubble rate and the deceleration parameter. Explicitly, it only appears in the second derivation of the Hubble rate\cite{ZP},
\begin{equation}\label{ddH}
\frac{\ddot{H}}{H^{3}} = \frac{9}{2} +
\frac{9}{2}w\frac{\rho_x}{\rho}\left[2 + w +
\frac{1}{3H}\left(\frac{Q}{\rho_{x}} -
\frac{\dot{w}}{w}\right)\right]\ .
\end{equation}
Its influence on the dynamics may be quantified by the
``statefinder parameter'' (``jerk'')
\begin{equation}\label{j}
j\equiv \frac{1}{aH^3}\frac{\mbox{d}^3 a}{\mbox{d}t^3} = 1 +
3\frac{\dot H}{H^2} + \frac{\ddot H}{H^3}\ .
\end{equation}
This parameter enters the luminosity distance
\begin{equation}\label{dL}
d_L =
\left(1+z\right)\int \frac{\mbox{d}z}{H\left(z\right)}
\end{equation}
in third order in the redshift:
\begin{eqnarray}\label{dL3}
d_L &\approx & \frac{z}{H_{0}} \left[1 + \frac{1}{2}\left(1 -q_0
\right)z \right.\nonumber\\
&&\left.+ \frac{1}{6}\left( 3\left(q_0 + 1\right)^2 - 5\left(q_0
+ 1\right) + 1 - j_0 \right)z^2\right].
\end{eqnarray}
While the $\Lambda$CDM model has $j = j_{0} = 1$, any interacting model has $j_{0} \neq 1$ in general,
which allows a discrimination between models that share the same values of $H_{0}$ e $q_{0}$.

\subsection{Scaling cosmology}

Starting point is an ansatz for the ratio of the energy densities of DM and DE,
\begin{equation}\label{rxi}
r = \frac{\rho_m}{\rho_x} = r_{0}a^{-\xi}
\end{equation}
with a phenomenological parameter $\xi$. The present value of the scale factor has been set to
 $a_{0}=1$. The $\Lambda$CDM model has $\xi = 3$.
A stationary ratio is characterized by $\xi =  0$.
The severity of the coincidence problem is quantified by $\xi$. According to Ref.~\onlinecite{dalal}, any $\xi < 3$
alleviates the coincidence problem.
It has been shown \cite{ZP}, that an interaction
\begin{equation}\label{Qscale}
Q = - 3 H\, \frac{\frac{\xi }{3}
+ w} {1 + r_{0} \left(1+z \right)^{\xi }}\, \rho _{m}\
\end{equation}
can produce any scaling behavior of the type (\ref{rxi}). The interesting special case of a stationary ratio
$r = r_0 = {\rm constant}$ results in a power-law solution
$\rho _{x},  \rho_{m} \propto a ^{-\nu }$, where  $\nu = 3 \frac{1 + r_0 + w}{1 + r_0}$.
This solution is able to describe an accelerated expansion under the condition  $3w < - \left(1+r_0\right)$. As any power-law solution, the dynamics can be described in terms of a scalar field with an exponential potential \cite{ZPC}.
The interesting feature here is that a specific interaction can give rise to a stationary ratio of the
energy densities which is consistent with an accelerated expansion of the Universe.

For a toy model $w = - 1$, $\xi = 1$ the impact of the interaction on observational quantities can be visualized as follows.
The total energy density of this model is
\begin{equation}\label{rhotoy}
\rho_m + \rho_x = \rho  = \frac{\rho _{0}}{\left(1+r_{0}
\right)^3}\left[1 + r_{0}a^{-1}\right]^{3}\ ,
\end{equation}
while one has
\begin{equation}\label{rhoLCDM}
\rho_{(\Lambda CDM) }  = \frac{\rho _{0}}{\left(1+r_{0} \right)}
\left[1 + r_{0}a^{-3}\right]\
\end{equation}
for the $\Lambda$CDM model.
The sum of powers in (\ref{rhoLCDM}) is replaced here by the power of a sum in (\ref{rhotoy}). This difference affects the luminosity distance which, for our toy model becomes
\begin{equation}\label{dtoy}
d_{\rm L} \approx \frac{z}{H_{0}} \left[1 + \frac{1 +
\frac{r_{0}}{4}}{1 + r_{0}}z  -
\frac{1}{8}\frac{r_{0}}{\left(1+r_{0}\right)^2}\left(6 +
r_{0}\right)z^2\right] \ ,
\end{equation}
in contrast to its $\Lambda$CDM counterpart
\begin{equation}\label{dlcdm}
d_{\rm L}^{\Lambda CDM} \approx \frac{z}{H_{0}}\left[1 + \frac{1 +
\frac{r_{0}}{4}}{1 + r_{0}}z  -
\frac{1}{8}\frac{r_{0}}{\left(1+r_{0}\right)^2}\left(10 +
r_{0}\right)z^2\right]\ .
\end{equation}
Both expressions differ in third order in the redshift.

What one would like to have is a dynamical evolution toward a constant ratio $r$. It has been shown in Ref.~\onlinecite{chimjak} that
indeed a suitable interaction between DM and DE can drive the
transition from a decelerated, matter dominated expansion to a
DE dominated period with accelerated expansion and a
stable, stationary ratio of the energy densities.

\subsection{Holographic dark energy}

\begin{figure}[!htb]
 \begin{minipage}[t]{0.70\linewidth}
\includegraphics[width=\linewidth]{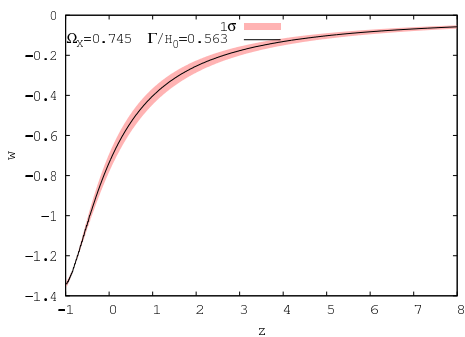}
\end{minipage} \hfill
    \caption{Evolution of the equation of state parameter of dark energy
    for the best fit model\cite{ivan}. The red swath indicates the
    region obtained by including the $1\sigma$ uncertainties of
    the constrained parameters
    $\Omega_{x}$ and $\Gamma/H_{0}$.} \label{fig:w(z)}
\end{figure}

An interaction modifies the $a^{-3}$ behavior of the
matter energy density to
\begin{equation}\label{rhommod}
\rho_{m} =
\rho_{m0}a^{-3}\,f(a)\ ,
\end{equation}
where the function $f(a)$ encodes the influence of the interaction.
The corresponding DE balance can be written in terms of an effective EoS parameter $w^{eff}$,
\begin{equation}\label{ebalhol}
\dot{\rho}_{x} = - 3H \left(1 + w^{eff}\right)\rho_{x}, \quad
w^{eff} = w + \frac{\dot{f}}{3 H f}\, r \ .
\end{equation}
In this context one can consider a constant energy density ratio:
\begin{equation}\label{rweff}
r = \mathrm{const} \  \Leftrightarrow\  w^{eff} = -
\frac{\dot{f}}{3 H f}\  \Rightarrow\  w^{eff} = \frac{w}{1 +
r}\ .
\end{equation}
The effective EoS coincides with the total EoS $\frac{p}{\rho} = w^{eff}$.
Via Friedmann's equation
\begin{equation}\label{friedmannhol}
3\,H^{2} = 8\,\pi\,G\,\left(\rho_{m} + \rho_{x}\right) =
8\,\pi\,G\,\left(1 + r\right)\rho_{x}
\end{equation}
it follows that for $r = \mathrm{const}$ we have necessarily $\rho_{x}\propto H^{2}$. Now the question arises, whether there exist DE models with $\rho_{x} \propto H^{2}$. This leads us to  holographic DE models.
Holographic DE models are characterized by an infrared cutoff length $L$ which is related
to an ultraviolet cutoff, corresponding to the DE density. As a consequence one obtains a dependence
$\rho_{x} \propto L^{-2}$.
The underlying idea here is that the energy in a volume $L^{3}$
should not exceed the energy of a black hole of the same
size \cite{cohen}:
\begin{equation}\label{bh}
\rho_{x}L^{3} \leq \frac{L}{8\,\pi\,G} \  \Rightarrow \
\rho_{x} = \frac{3\,c^{2}}{8\,\pi G\,L^{2}}\ ,
\end{equation}
where $c^{2}$ is a constant of the order of unity.
A choice $L = H^{-1}$ then provides us with
\begin{equation}\label{rhohol}
\rho_{x}
= \frac{3\,c^{2}\,H^{2}}{8\,\pi\,G}\ .
\end{equation}
This expression was shown \cite{cohen} to yield a correct order of magnitude of the presently observed value of the cosmological term.
From (\ref{rweff}) it follows that under this condition $w^{eff}$ and $w$ are only different from zero for a non-vanishing interaction. In the non-interacting limit the only possible EoS is $w=0$, which is incompatible with an accelerating universe. A negative EoS parameter is a pure interaction effect. This becomes particularly clear if we look at the expression
\begin{equation}\label{qhol}
q = \frac{1}{2}\left(1 - \frac{\dot{f}}{H f}\right)
\end{equation}
for the deceleration parameter $q$.
Obviously, the sign of $q$ depends on the ratio $\frac{\dot{f}}{H f}$.
For $\frac{\dot{f}}{f} < H$ we have $q > 0$, while $q < 0$ holds for $\frac{\dot{f}}{f} > H$.
The interesting question is, whether an
evolution from $\frac{\dot{f}}{f} < H$ to  $\frac{\dot{f}}{f}
> H $ is possible. This would be equivalent to a transition from decelerated to accelerated expansion as a pure interaction phenomenon. Let us consider to this purpose the simple example of a constant interaction rate $\Gamma$ where $\Gamma = \frac{Q}{\rho_{x}}$. In this case, the
Hubble rate is \cite{zpCQG}
\begin{equation}\label{Hhol}
H = H_{0}\left[\frac{\Gamma}{3 H_{0}r} + \left(1 -
\frac{\Gamma}{3 H_{0}r}\right)a^{-3/2}\right]\ ,
\end{equation}
which represents a special generalized Chaplygin gas. The evolution of the EoS parameter is shown in Fig. \ref{fig:w(z)}, which confirms that there is indeed a transition from $w \approx 0 $ at high redshifts to an EoS parameter $w \approx -1$ at present.

\begin{figure}[!htb]
\begin{minipage}[t]{0.70\linewidth}
\includegraphics[width=\linewidth]{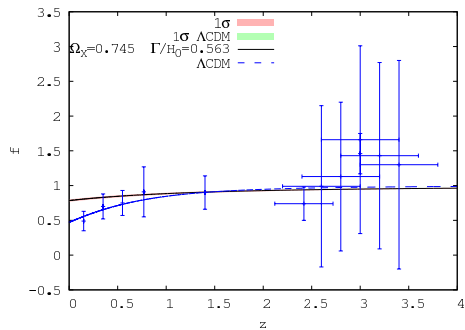}
\end{minipage} \hfill
    \caption{Growth function vs. redshift for the best fit holographic model
    (solid line)\cite{ivan}. Also shown is the prediction of the $\Lambda$CDM model (dashed
    line).}
    \label{fig:f(z)}
\end{figure}

Introducing the density contrast $\delta_{m} = \frac{\hat{\rho}_{m}}{\rho_{m}}$, where $\hat{\rho}_{m}$ is the first-order DM density perturbation and assuming for simplicity that the corresponding DE perturbations $\delta_{x} = \frac{\hat{\rho}_{x}}{\rho_{x}}$ are related to the matter perturbations by $\delta_{x} = \alpha
\delta_{m}$ with a constant $\alpha$, we obtain the following closed equation for the DM density contrast\cite{zpCQG}:
\begin{eqnarray}\label{delta2pr}
\delta_{m}^{\prime\prime} &+& \frac{3}{2a}\left[1 +
\frac{\Gamma}{3Hr} +
\frac{2\left(1-\alpha\right)}{3}\frac{\Gamma}{Hr}\right]\delta_{m}^{\prime}\nonumber \\
&-&\frac{3}{2a^{2}}\frac{r+\alpha}{r+1}\left[1 -
\frac{4\left(1-\alpha\right)}{3}
\frac{\Gamma}{Hr}\frac{r+1}{r+\alpha}\right]\delta_{m} = 0\ .\nonumber\\
\end{eqnarray}
The prime denotes derivation with respect to the scale factor.
Alternatively, the fluctuation dynamics is characterized by the
growth rate parameter $f\equiv \frac{d \ln \delta_{m}}{d \ln a}$, which obeys the equation
\begin{equation}\label{eqf}
f^{\prime}\,  + \, f^{2} + \, \frac{1}{2}\left(1 +
\frac{\Gamma}{Hr}\right)f - \frac{3}{2} = 0\ .
\end{equation}
In the limit $\Gamma = 0$ we have $f = 1$, i.e., we recover the Einstein-de Sitter behavior
$\delta_{m}\propto a$. For the model based on (\ref{Hhol}) and (\ref{eqf}), the quantity $f$ is visualized and compared with observational data in Fig.~\ref{fig:f(z)}.



\subsection{Transient acceleration}

\begin{figure}[!t]
\begin{minipage}[t]{0.70\linewidth}
\includegraphics[width=\linewidth]{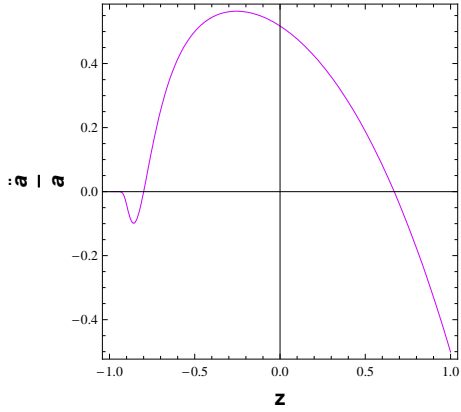}
\end{minipage} \hfill
\caption{\label{2d}{\protect\footnotesize Acceleration parameter for the best fit scenario
in units of $H_{0}^{2}$ (taken from Ref.~\onlinecite{nelson}).
}}
\label{transient}
\end{figure}

There are results in the literature which suggest that cosmic acceleration could be a transient phenomenon \cite{shaf}.
A dynamics of this type may be the result of an interaction between DM and DE as can be shown by the toy model considered in Ref.~\onlinecite{nelson}.
Let us parametrize the interaction according to
\begin{equation}\label{fg}
f\left(a\right)=1+g\left(a\right)
\end{equation}
and consider the special case
\begin{equation}\label{gauss}
w= -1 \ , \quad g(a)=\gamma\, a^5 \exp (-a^2/\sigma ^2)\ ,
\end{equation}
where $\gamma$ is an interaction constant.
The DE density becomes
\begin{equation}\label{rhoxtrans}
\rho_{x} = \rho_{x_{0}}^{eff}
- \gamma\,\frac{\rho_{m_0}}{1+g_{0}}\,\exp\left(-a^{2}/\sigma^{2}\right) \left(a^{2} - \frac{3}{2}\sigma^{2}\right)
\end{equation}
with an effective cosmological constant
\begin{equation}\label{effconst}
\rho_{x_{0}}^{eff}
= \rho_{x_{0}} - \frac{3}{2}\gamma\,\frac{\rho_{m_0}}{1+g_{0}}\,\exp (-1/\sigma^{2})
\left[\sigma^{2} - \frac{2}{3}\right]\ .
\end{equation}
The interaction renormalizes the bare (interaction-free) value $\rho_{x_{0}}$.
A transient acceleration is only possible for $\rho_{x_{0}}^{eff}=0$ since otherwise the constant
would always prevail in the long-time limit. This means, for accelerated expansion to be a transient phenomenon, part of the interaction has to cancel the bare cosmological constant.
Under this condition and with
\begin{equation}\label{defK}
 K = \frac{8 \pi G}{3 H_{0}^{2}}\gamma\frac{\rho_{m_0}}{1+g_{0}}
\end{equation}
we obtain
\begin{eqnarray}
\frac{\ddot a}{a}&=& -\frac{1}{2}H_{0}^{2}
\left\{\frac{1 - \frac{3}{2}K\sigma^{2}
\exp
(-1/\sigma^{2})}{a^{3}}\right\}\\
 &-&\frac{1}{2}H_{0}^{2}
\left\{-
3 K \exp
(-a^{2}/\sigma^{2}) \left[\sigma^{2} -a^{2}\right]\right\}\ .
\label{ddatrans}
\end{eqnarray}
This may be compared with the corresponding expression for the $\Lambda$CDM model
\begin{equation}\label{ddalcdm}
\frac{\ddot a}{a}=-\frac{1}{2}H_{0}^{2}
\left\{\frac{1 - \Omega_{\Lambda}}{a^{3}}
- 2 \Omega_{\Lambda}\right\}\ .
\end{equation}
Since one expects an alternative model not to deviate too strongly from the $\Lambda$CDM model at the present time, the comparison between (\ref{ddatrans}) and (\ref{ddalcdm}) suggests $K$ to be positive and the interaction term to play a similar role as the cosmological constant $\Lambda$. In other words, the role of the interaction is twofold. As already mentioned, it has to cancel the bare cosmological constant. But at the same time it has to induce an accelerated expansion by itself.
The expression (\ref{ddatrans})  guarantees an
early ($a\ll 1$) decelerated expansion for $\Omega_{m_{0}} > K\exp (-1/\sigma^{2})$ which represents an
upper limit on the interaction strength $K$.
To reproduce the presently observed accelerated expansion
requires the condition
\begin{equation}\label{K>}
\frac{\ddot{a}}{aH^{2}}\mid _{0} \ > 0 \quad \Leftrightarrow \quad
K\,\exp (-1/\sigma^{2})\left[\sigma^{2} - \frac{2}{3}\right] > \frac{2}{9}\ ,
\end{equation}
which represents a lower limit on the interaction strength.
It follows that there is an admissible
range
\begin{equation}\label{range}
\frac{2}{9}\frac{e^{1/\sigma^{2}}}{\sigma^{2} - \frac{2}{3}} < K <
\frac{2 e^{1/\sigma^{2}}}{3\sigma^{2}}
\end{equation}
for $K$.
From the data of the GOLD sample we found best-fit values $\sigma = 5.000$, $K = 0.019$, $h = 0.644$,  $\Omega_{x0} = 0.678$ and $\chi^2_{min} = 1.129$ which have to be contrasted with the corresponding values for the $\Lambda$CDM model which are $h = 0.64$,  $\Omega_{x0} = \Omega_{\Lambda0} = 0.65$ and $\chi^2_{min} = 1.128$. Fig.~\ref{transient} shows that this model indeed describes an early transition
from decelerated to accelerated expansion together with a future transition back to decelerated expansion\cite{nelson}.

\subsection{Decaying vacuum energy}

A still alternative scenario is based on a prescribed behavior of the decay of the cosmological term, interpreted as vacuum energy.
The only preferred time scale in a homogeneous and isotropic universe is the Hubble time
$H^{-1}$. It is tempting to associate a supposed vacuum decay with this scale.
The simplest case therefore is a dependence $\rho_{X} \propto H$. This relation has also some support from QCD \cite{QCD}, but our approach here is phenomenological.
The dark-energy density is characterized by
\begin{equation}\label{rhodecvac}
\rho_{x} = \frac{\sigma}{3}\Theta\ ,\ p_{x} = - \rho_{x}\ ,\   \Theta \equiv u^{a}_{;a}
\end{equation}
which, in the homogeneous and isotropic background reduces to $\rho_{x} = \sigma H$,
$p_{x} = - \sigma H$ and $\Theta = 3H$.
For the Hubble rate we find
\begin{equation}\label{Hdec}
H = H_{0}\left[1 - \Omega_{m0} + \Omega_{m0}a^{-3/2}\right]\ ,
\end{equation}
where we used that $\sigma = \frac{\rho_{0}}{H_{0}}\left(1 - \Omega_{m0}\right)$.
The ratio of the energy densities scales as
\begin{equation}\label{ratiodec}
\frac{\rho_{m}}{\rho_{x}} = \frac{\Omega_{m0}}{1 - \Omega_{m0}}a^{-3/2}\ .
\end{equation}

To consider the perturbation dynamics we split the interaction term $Q^{i}$ in (\ref{coupling}) - (\ref{mbalx})
into parts parallel and perpendicular to $u^{i}$,
\begin{equation}\label{Qsplit}
Q^{i} = u^{i}Q + \bar{Q}^{i}\ ,
\end{equation}
such that
\begin{equation}\label{Qsplit2}
Q = - u_{i}Q^{i}\ ,\  \bar{Q}^{i} = h^{i}_{a}Q^{a}\ , \  u_{i}\bar{Q}^{i} = 0\ .
\end{equation}
The basic set of equations then is
\begin{equation}\label{mbaldec}
\dot{\rho}_{m} + \Theta \rho_{m} = -\frac{\sigma}{3}\dot{\Theta}\ , \ \
\rho_{m}\dot{u}_{a} = \frac{\sigma}{3}h_{a}^{i}\Theta_{,i}
\end{equation}
and the Raychaudhuri equation
\begin{equation}\label{Ray}
\dot{\Theta} + \frac{1}{3}\Theta^{2} - \dot{u}^{a}_{;a} + 4\pi G \left(\rho + 3
p\right) = 0\ .
\end{equation}
The line element for scalar perturbations is written as
\begin{eqnarray}\label{line}
\mbox{d}s^{2} &=& - \left(1 + 2 \phi\right)\mbox{d}t^2 + 2 a^2
F_{,\alpha }\mbox{d}t\mbox{d}x^{\alpha} \nonumber\\ && +
a^2\left[\left(1-2\psi\right)\delta _{\alpha \beta} + 2E_{,\alpha
\beta} \right] \mbox{d}x^\alpha\mbox{d}x^\beta\ .
\end{eqnarray}
For the spatial components of the four-velocity we write
\begin{equation}\label{v}
a^2\hat{u}^\mu + a^2F_{,\mu} = \hat{u}_\mu \equiv v_{,\mu}\ ,
\end{equation}
thus introducing the velocity potential $v$.
The first-order matter energy balance becomes
\begin{equation}\label{mbal1st}
\dot{\delta}_{m} + 4\pi G \sigma \delta_{m} - \phi \left(- 3H + 4\pi G\sigma\right) + \hat{\Theta} = \frac{\hat{Q}}{\rho_{m}}\ .
\end{equation}
Now it is convenient to formulate the perturbation dynamics in terms of the
gauge-invariant quantities
\begin{equation}\label{invc}
\hat{\Theta}^{c} \equiv \hat{\Theta} + \dot{\Theta} v \ , \  \delta_{m}^{c} \equiv \delta_{m} + \frac{\dot{\rho}_{m}}{\rho_{m}} v \ ,\  \delta_{x}^{c} \equiv \delta_{x} + \frac{\dot{\rho}_{x}}{\rho_{x}} v\ .
\end{equation}
Then the energy- and momentum balances are combined into
\begin{equation}\label{balcomb}
\dot{\delta}_{m}^{c} + \frac{8\pi G}{3} \sigma \delta_{m}^{c}  + J\hat{\Theta}^{c} = 0\ ,
\end{equation}
where
\begin{equation}\label{J}
J \equiv 1 + \frac{\sigma H}{3\rho_{m}}\left(1 - \frac{4\pi G}{3}\frac{\sigma}{H} - \frac{\sigma H}{3\rho_{m}}\,\frac{k^{2}}{a^{2}H^{2}}\right)
\end{equation}
is a scale-dependent ($k$ denotes the comoving wavenumber) quantity.
Differentiation of (\ref{balcomb}) and combination with the first-order Raychaudhuri equation for $\hat{\Theta}^{c}$
yields the second-order equation \cite{saulo}
\begin{equation}\label{deltamfin}
\delta_m^{c\prime\prime} + g(a,k)\delta_m^{c\prime} + f(a,k)\delta_m^{c} = 0\ ,
\end{equation}
where the coefficients $g(a,k)$ and $f(a,k)$ are known, scale-dependent background functions (see Ref.~\onlinecite{saulo}).
It is an advantage of this model that it allows us to calculate the dark-energy perturbations in terms of the matter perturbations and their first derivative. The relevant relation is
\begin{equation}\label{deltaxdec}
\delta_{x}^{c} = - \frac{1}{3J}\left[a\delta_{m}^{c\prime}
+ \frac{1 - \Omega_{M0}}{1 - \Omega_{M0} + \Omega_{M0}a^{-3/2}}\delta_{m}^{c}\right]\ .
\end{equation}
From the definition (\ref{J}) of $J$ it follows that on small scales
$\frac{k^{2}}{a^{2}} \gg H^{2}$ one has $|J|\gg 1$, equivalent to
$\delta_{x}^{c} \ll \delta_{m}^{c}$, i.e., the
DE perturbations are negligible on these scales.

Pressure perturbations in interacting two-component models are generally non-adiabatic.
The relevant quantity is
\begin{equation}\label{hatp}
\hat{p}- \frac{\dot{p}}{\dot{\rho}}\hat{\rho} =  \frac{\dot{\rho}_{x}\dot{\rho}_{m}}{\dot{\rho}}
\left(\frac{\hat{\rho}_{m}}{\dot{\rho}_{m}} - \frac{\hat{\rho}_{x}}{\dot{\rho}_{x}}\right)\ .
\end{equation}
For adiabatic perturbations
$\hat{p}- \frac{\dot{p}}{\dot{\rho}}\hat{\rho} = 0$ is valid.

At high redshift, equation (\ref{deltamfin}) reduces to
\begin{equation}\label{deltaeds}
\delta^{c\prime\prime}_{m} + \frac{3}{2a} \,\delta^{c\prime}_{m} - \frac{3}{2a^2}\,\delta^{c}_m =0 \  \qquad \qquad  (a \ll 1) \ ,
\end{equation}
which coincides with the corresponding equation for an Einstein-de Sitter universe. 
The non-adiabaticity is very small for $a \ll 1$ but it vanishes exactly only for $\Omega_{m0}=1$. Under this condition it is possible to impose adiabatic initial conditions for the evolution of the matter perturbations which are then used to calculate the matter power spectrum.
\begin{figure}[t!]
\begin{minipage}[t]{0.60\linewidth}
\includegraphics[width=\linewidth]{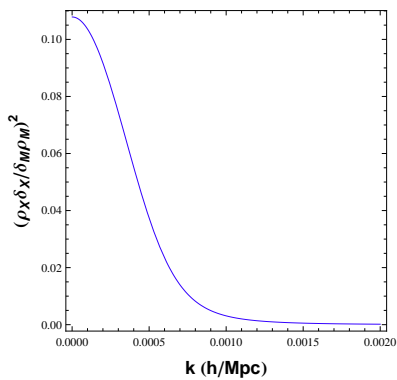}
\end{minipage}
\caption{{\protect\footnotesize Relative power spectrum as a function of $k$ on large scales for $\Omega_{m0} = 0.3$ (taken from Ref.~\onlinecite{saulo}).}}
\label{LS}
\end{figure}
This corresponds to the circumstance that DE perturbation are very small for small values of the scale factor,
\begin{equation}\label{fracrx}
\frac{|\hat{\rho}_{x}|}{|\hat{\rho}_{m}|}  \approx \frac{1}{3} \frac{1 - \Omega_{m0}}{\Omega_{m0}}a^{3/2}\qquad \qquad  (a \ll 1)\ .
\end{equation}

The relative power-spectrum, based on equations (\ref{deltamfin}) and (\ref{deltaxdec}), is shown in Fig.~\ref{LS}. It indicates an increase of DE perturbations on very large scales.
\\

\section{Summary}
We have summarized here various aspects of the potential role of interactions in the cosmological dark sector. In the simplest case this comprises third-order redshift corrections in the luminosity distances of supernovae of type Ia. But a suitable coupling may also replace the cosmological constant and generate a phase of accelerated expansion by itself. This can happen, e.g., in the context of holographic DE models and in models of transient acceleration. In the latter case the future evolution of the Universe is qualitatively different from that predicted by the $\Lambda$CDM model. A decaying-vacuum-energy approach is another scenario in which interactions have a noticeable impact on the cosmological
dynamics. Quite generally, interacting models are characterized by non-adiabatic pressure perturbations, even though the individual components are adiabatic on their own.

\begin{acknowledgments}
Partial support by CNPq (Brazil) is gratefully acknowledged.
\end{acknowledgments}


\end{document}